\documentclass[a4paper,twoside,11pt]{article}
\usepackage{amssymb,amsfonts,amstex,axodraw}
\usepackage{latexsym}
\usepackage{newcent}
\usepackage{umlaut}
\begin{document}

\title{The Hopf Algebra Structure of Connes and Kreimer 
in Epstein-Glaser Renormalization}

\author{G. Pinter}

\maketitle

\vspace{-1.1cm}
\begin{center}
\textsc{%
II. Institut f\"ur Theoretische Physik\\
Universit\"at Hamburg\\
Luruper Chaussee 149\\
22761 Hamburg 
Germany}\\
e-mail: \verb+gudrun.pinter@desy.de +
\end{center}
 
\vspace{1cm}
\begin{abstract}
\noindent
We show how the Hopf algebra structure of renormalization discovered 
by Kreimer can be found in the Epstein-Glaser framework without using
an analogue of the forest formula of Zimmermann.
\end{abstract}

\section{Introduction}

Epstein-Glaser renormalization is the mathematical formulation
of what is done in renormalization theory. It is formulated
in position space and its applications up to now are of more abstract
nature than explicit calculations, which are up to now
predominantly made in momentum space.
The combinatoric of momentum space renormalization
is described by Bogoliubovs recursion formula
or equivalently by the forest formula of Zimmermann.
Recently Kreimer discovered a Hopf algebra structure in the 
combinatorics of the subtraction procedure in momentum space.
Connes and Kreimer worked out the connection of this algebra
to the Riemann-Hilbert problem \cite{rmh} and to noncommutative
geometry \cite{ncg}. For a survey of the present status and the literature
in this field see \cite{kreineu}.
In this work we derive the structure of this Hopf algebra
in Epstein Glaser renormalization in a completely different way than
Bond\'{\i}a and Lazzarini \cite{laz} recently did. 
A short version of the
derivation given here is presented in \cite{doc}.

The main difficulty to transfer the work of Connes and Kreimer      
into the Epstein-Glaser framework
is that the recursive definition of $T$-products
gives no insight into the combinatorics.
Furthermore there is a different definition
of subdiagrams in position space.
Bond\'{\i}a and Lazzarini \cite{laz}
discussed the Hopf algebra
structure in Epstein Glaser renormalization, but they 
adapted a formulation of the forest formula of Zimmermann
with the following form:

\begin{eqnarray}
R_{\Gamma} f(\Gamma) = \left[ 1 + \sum_{\mathcal{F}} \prod_{
\gamma \in \mathcal{F}} (-S_{\gamma}) \right] f(\Gamma), 
\end{eqnarray}

where the sum is over all nonempty sets whose elements are
proper, divergent (and may be connected) subgraphs of
$\Gamma$. 
The problem with this formula is that
$R_{\Gamma}$ is not an operator defined on the space of
distributions. Especially it cannot be adjoint to give
an operator on the testfunction $f(\Gamma)$ is smeared with.

Starting point of our derivation is the structure of finite
renormalizations proved in \cite{ich}. We show that the 
divergencies of the numerical distributions in Epstein-Glaser
renormalization posess a Hopf algebra structure 
analogously to the structure described in \cite{rmh}.

Therefore we describe in the first section
the Hopf algebra of graphs and
subgraphs, which is different from the one in \cite{rmh}
because of the different definition of subgraphs.
Then we recall the structure of finite renormalizations
and show that the singularities of the numerical distributions
obey the same equation as the antipode of the Hopf algebra
of the corresponding algebra of graphs.

\section{The Hopf Algebra of Graphs}

We take the algebra $\mathcal{A}$ of graphs with multiplication
and addition as in \cite{rmh}, but with connected instead of
one particle irreducible graphs:

$\mathcal{A}$ has a basis labeled by all Feynman graphs $\Gamma$
which are disjoint unions of connected graphs:

\begin{eqnarray}
\Gamma = \bigcup _{j=1}^n \Gamma_j
\end{eqnarray}

The empty graph $\Gamma = \emptyset$ is the unit element 
$\mathbb{I}$ of the algebra.

The product in $\mathcal{A}$ is bilinear and defined on the basis
by the operation of disjoint union:

\begin{eqnarray}
m: \mathcal{A} \otimes \mathcal{A} & \rightarrow & \mathcal{A} 
\nonumber \\
m(\Gamma_1,\Gamma_2 )=
\Gamma_1 \cdot\Gamma_2 & = & \Gamma_1 \cup \Gamma_2
\end{eqnarray}

This means that the product of two graphs is a graph consisting
of this two disconnected subgraphs.

For the definition of a coproduct we need the notion of a subgraph
in Epstein-Glaser renormalization.
In a graph $\Gamma$ with $n$ vertices labeled with the set of
indices $J=\{ 1, \ldots, n\}$ each subset $I$ of $J$ defines a
subgraph $\gamma _I$ consisting of all vertices labeled by $I$
and all lines joining them.

In momentum space, used in \cite{rmh}, a subgraph is defined in contrast
by the lines belonging to it.
All vertices which are endpoints of lines of a subgraph belong
to the subgraph. 
So in momentum space the set of subgraphs is larger because not all 
lines between two vertices of the graph have to belong to the subgraph.

If the set $I$ defining a subgraph 
consists of only one element, the subgraph has only
one point and is defined as the empty graph $\mathbb{I}$.

Every partition $P$ of $J$ defines a set of subgraphs, more 
precisely if $P=\{ O_1, \ldots, O_k\}$ the subgraphs are those
belonging to the subsets $O_i$. We use the following notation:

\begin{eqnarray}
\gamma_P = \{ \gamma_{O_1}, \ldots, \gamma_{O_k} \}.
\end{eqnarray}

$\Gamma \setminus \gamma _P$ is obtained from
$\Gamma$ by shrinking its nonempty connected
subgraphs belonging to
$\gamma_P$ to a point.

Following
\cite{rmh} we restrict ourselves to a scalar massive theory.
In this case we have two two-line vertices according to the terms
$\Phi^2$ and $\partial_{\mu} \Phi \partial^{\mu} \Phi$ in the Lagrangian.

Now we define a coproduct on $\mathcal{A}$. For a 
connected nonempty graph
$\Gamma$ with $|J|$ vertices we define:

\begin{eqnarray}
c & : &\mathcal{A} \to \mathcal{A} \otimes \mathcal{A} \nonumber \\
c(\Gamma)& =& \Gamma \otimes \mathbb{I} + \mathbb{I} \otimes \Gamma +
\sum_{P \in Part J \atop 1 \not= |P| \not= n} \nolimits ' 
\gamma_P^{(i)} \otimes \Gamma \setminus
\gamma_P^{(i)} .
\end{eqnarray}

where the sum $ \sum'$ is over all partitions with subgraphs  
$\gamma_{O_i}$ where $\gamma_{O_i}$ is the empty graph or a graph
with a superficial divergence (in $\Phi ^4$ theory all graphs
with only two or four external lines).
The multiindex $(i)$ has one value for each connected component of
$\gamma_P$. This value is 0 or 1 for a component with two
external lines and 0
for all other components, corresponding to the two different
two-line vertices.

In the special case $\Gamma = \emptyset$ we define

\begin{eqnarray}
c(\emptyset) = \mathbb{I} \otimes \mathbb{I}.
\end{eqnarray}

Since this definition varies from the one given in \cite{rmh} only by the
definition of subgraphs,
the proof of the coassociativity is analogous to that of \cite{rmh},
if we substitute 1PI graphs by connected graphs.

A counit is defined by 

\begin{eqnarray}
\bar{e}(\Gamma) &  = & 1 \qquad \mbox{for} \qquad \Gamma = 
\emptyset \nonumber \\
\bar{e}(\Gamma) & = & 0 \qquad \mbox{for} \qquad \Gamma \not= \emptyset
\end{eqnarray}

The defining equation for the antipode $s: \mathcal{A} \to
\mathcal{A}$ is

\begin{eqnarray}
m ( s \otimes id) c(\Gamma) = \bar{e}(\Gamma) \cdot \mathbb{I}.
\end{eqnarray}

($s$ is an antihomomorphism of algebras).
From this equation we obtain a recursion formula for the
antipode:

\begin{eqnarray}
s(\Gamma) = - \Gamma - \sum_{P \in Part(J) \atop 1 \not= |P|
\not= n} \nolimits '
s( \gamma _P^{(i)}) \cdot \Gamma \setminus  \gamma_P^{(i)}
\label{antip}
\end{eqnarray}

for $\gamma \not= \mathbb{I}$ and

\begin{eqnarray}
s(\mathbb{I})= \mathbb{I},
\end{eqnarray}

where $s(\gamma_P^{(i)})= \prod _{O_j \in P } s(\gamma_{O_j}^{(i_j)})$,
where $(i_j)$ is the part of the multiindex $(i)$ belonging to $O_j$.

We now show that we can derive an equation 
of the same form for
the singularities of the numerical distributions
belonging to these graphs.

\section{Structure of Epstein-Glaser Renormalization}

The main objects studied
in Epstein-Glaser renormalization are $T$-products. 
Usually $T$-products are defined as multilinear functions on 
Wick monomials of quantized fields \cite{fredneu}. Here we use
another definition of $T$-products, which was first introduced in
\cite{fm}.
Let $\mathcal{A}$ be a commutative algebra generated by the
so-called classical symbolical 
fields $\phi_i$ and their derivatives. 
We regard

\begin{eqnarray}
\mathcal{D} \left( \mathbb{R}^4, \mathcal{A} \right) \ni f =
\sum_i g_i \phi _i, 
\end{eqnarray}

where the sum is a finite sum over elements $\phi_i$ of the
algebra $\mathcal{A}$.
An element $f$ is then given by its coefficients $g_i \in \mathcal{C}
^{\infty}_0 (\mathbb{R}^4)$.

The time-ordered product ($T$-product) is a family of maps
$T_n, n \in \mathbb{N}$, called $T_n$-products. They are functions from
$\left( \mathcal{D} \left( \mathbb{R}^4, \mathcal{A} \right) \right)
^{\otimes n}$
into the set of operators on the Hilbert space
$\mathcal{H}$ with the following properties:

\begin{enumerate}

\item $T_0 = 1$
 
$T_1 \left( f \right)= \sum_i :\phi _i (g_i): \qquad \forall \quad
f \in \mathcal{D} \left( \mathbb{R}^4, \mathcal{A} \right),$

where the sum is taken over all generators of $\mathcal{A}$.
Each local field is the image of an element 
$f \in \mathcal{D} \left( \mathbb{R}^4, \mathcal{A} \right)$ under $T_1$.

\item Symmetry in the arguments:  

\begin{eqnarray}
T_n \left(  f_1, \ldots ,f_n \right) =
T_n \left(  f_{\pi_1},  \ldots , f_{\pi_n} \right) 
 \qquad \forall \  \pi \in S_n \qquad 
\forall \ f_i \in \mathcal{D} \left( \mathbb{R}^4, \mathcal{A} \right),
i=1, \ldots n, \nonumber 
\end{eqnarray}

where $S_n$ is the set of all permutations of $n$ elements.

\item The factorization property:

\begin{eqnarray}
T_n \left(  f_1,  \ldots ,f_n  \right) =
T_i \left(  f_1, \ldots ,f_i \right)
T_{n-i} \left(  f_{i+1}, \ldots ,f_n \right) \label{fac}
\end{eqnarray}

if \ \
$(\mathrm{supp} f_1 \cup \ldots \cup \mathrm{supp} f_i) \ \gtrsim \
(\mathrm{supp} f_{i+1} \cup \ldots \cup \mathrm{supp} f_n) \ $
and \
$f_i \in \mathcal{D} \left( \mathbb{R}^4, \mathcal{A} \right) \ \forall \ i$.

\end{enumerate}

The $T$-products are not uniquely determined by these properties,
so there are different $T$-products. Popineau and Stora \cite{stora}
proved that there exist finite renormalizations between two 
different $T$-products.
We know from \cite{ich} that 
the combinatorics in Epstein-Glaser renormalization are hidden
in the structure of finite renormalizations (differences between
two different $T$-products). 
For two different $T$-products denoted by $T$ and $\hat{T}$ there
exist functions $\Delta_n : \mathcal{D} \left( \mathbb{R}^{4n}
,\mathcal{A}^n \right) \to \mathcal{D} \left( \mathbb{R}^{4},\mathcal{A} 
\right)$
with $\mathrm{supp} \ \Delta_n \subset D_n$ and

\begin{eqnarray}
\hat{T}_n \left( \mbox{$\bigotimes$} _{j \in J} f_j \right) =
\sum_{ P \in Part(J) }
T_{|P|} \Bigl[ \bigotimes_{O_i \in P}
\Delta_{|O_i|} \left( \mbox{$\bigotimes$}_{j \in O_i} f_j \right)
\Bigr]. \label{nkl}
\end{eqnarray}

In the proof of this equation
given in \cite{ich} the $\Delta_n$ are constructed
inductively by the following formula:

\begin{eqnarray}
T_1 \left( \Delta_n \left( \mbox{$\bigotimes$} _{j \in J} f_j 
\right)\right) & = & 
\hat{T}_n \left( \mbox{$\bigotimes$} _{j \in J} f_j \right)
- \sum_{ P \in Part(J) \atop |P|>1} T_{|P|} \Bigl[ \bigotimes_{O_i \in P}
\Delta_{|O_i|} \left( \mbox{$\bigotimes$}_{j \in O_i}  f_j \right)
\Bigr] . \nonumber \\ \label{poi}
\end{eqnarray}

A $T_n$-product is a sum of terms each of which can be 
associated to the contribution of a special Feynman graph
with $n$ vertices.
We now only regard the contribution of $T_n$ to a special 
Feynman graph 
$\gamma$ with $n$ vertices, denoted by $T_n^{\gamma}$. 
With the notion of subgraphs explained in the previous section 
we obtain from (\ref{nkl}):

\begin{eqnarray}
\hat{T}_n ^{\gamma}
\left( \mbox{$\bigotimes$} _{j \in J} f_j \right) =
T_{n} ^{\gamma} \left( \mbox{$\bigotimes$} _{j \in J} f_j \right) +
\sum_{ P \in Part(J) \atop |P|<n  }
T_{|P|} ^{\gamma \setminus \gamma_P}
\Bigl[ \bigotimes_{O_i \in P}
\Delta_{|O_i|} ^{\gamma_{O_i}}
\left( \mbox{$\bigotimes$}_{j \in O_i} f_j \right)
\Bigr], \label{strepgl}
\end{eqnarray}

where $\Delta ^{\gamma}$ is defined inductively by

\begin{eqnarray}
T_1 \left( \Delta_n^{\gamma} \left( \mbox{$\bigotimes$} _{j \in J} f_j 
\right)\right) & = & 
\hat{T}_n^{\gamma} \left( \mbox{$\bigotimes$} _{j \in J} f_j \right)
- \sum_{ P \in Part(J) \atop |P|>1} T_{|P|}^{\gamma \setminus \gamma_P}
 \Bigl[ \bigotimes_{O_i \in P}
\Delta_{|O_i|}^{\gamma} \left( \mbox{$\bigotimes$}_{j \in O_i}  f_j \right)
\Bigr]. \nonumber \\
\end{eqnarray}

Comparing the recursive formula of renormalization of Bogoliubov
(which can be found in \cite{zim1}  for example)
with (\ref{strepgl}) we see that both 
renormalizations consist of a sum of terms, each belonging to a
set of superficially divergent parts of $\Gamma$. 
The superficially convergent parts do not contribute to the
sum in (\ref{strepgl}) because in this case $\Delta$ is 0, and we can
replace the sum $\sum$  by the sum $\sum '$ which is over all
partitions with empty subgraphs or subgraphs with a superficial
degree of divergence.

In contrast to Bogoliubovs 
formula equation (\ref{strepgl}) is formulated for finite 
renormalizations.

In the following we omit the lower indices of the 
$T$-products and their arguments $f$ and obtain with
$\Delta ^{\gamma_P} = \prod _{O_i \in P} \Delta ^{\gamma_{O_i}}$:

\begin{eqnarray}
\hat{T} ^{\Gamma} =
T ^{\Gamma} + \Delta^{\Gamma}+
\sum_{ P \in Part(J) \atop 1<|P|<n  } \nolimits '
T^{\Gamma \setminus \gamma_P}
\Bigl[ 
\Delta^{\gamma_P} 
\Bigr]. \label{gleichung} 
\end{eqnarray}

We now restrict ourselves to the case of $\Phi^4$ theory.
The graphs are built of one kind of line, two kinds of
two line vertices and one kind of four line vertex.
A generalization to other theories is obvious but
more complicated to write down.
In \cite{ich} it is shown that if $\gamma$ is a diagram
with four external legs, $\Delta ^{\gamma}$ has the form

\begin{eqnarray}
\Delta ^{\gamma} _n ((g \phi^4)^{\otimes n})
= C^{(n)}g^n (v) \phi^4,
\end{eqnarray}

where the constants
$C^{(n)}$ are distributions (more precise products of
Feynman propagators) smeared with testfunctions which are
derivatives of functions $w$ needed in the subtraction procedure.
In the case where $\gamma$ is the empty graph we have
$\Delta^{\gamma}(f) = f$ and if $\gamma$ has two external
lines we have

\begin{eqnarray}
\Delta ^{\gamma} _n =  A^{(n)}g^n (v) \phi^2
+B^{(n)}g^n (v) \partial ^{\mu}\phi
\partial _{\mu}\phi ,
\end{eqnarray}

where $A^{(n)}$ and $B^{(n)}$ have the same form as $C^{(n)}$.

Now we formulate equation (\ref{gleichung})
on the level of numerical distributions, this means we omit the
Wick products and the smearing functions.
According to the values of $\Delta^{\gamma_P}$ we denote with
$\delta ^{\gamma_P^{(i)}} = \prod_{O_j \in P} \delta^{\gamma_{O_j}
^{(i_j)}}$ the numerical distribution belonging to 
$\Delta ^{\gamma_P^{(i)}} = \prod_{O_j \in P} \Delta^{\gamma_{O_j}
^{(i_j)}}$.
As in the previous chapter the multiindex $(i)$ has a value for 
each connected component of $P$ and $(i_j)$ is the subset of
$(i)$ belonging to the connected components of $O_j$.
For a connected component of $P$ with two external legs the
value can be 0 or 1, if the value is 0 we take in the product
the numerical distribution of the coefficient $A^{(n)}$
and if the value is 1 we take the numerical distribution of the
coefficient $B^{(n)}$.
With 
$t^{\gamma}$ we denote the numerical distributions 
belonging to the graphs $\gamma$.
We obtain

\begin{eqnarray}
\hat{t} ^{\Gamma} =
t ^{\Gamma} + \delta^{\Gamma} +
\sum_{ P \in Part(J) \atop 1<|P|<n  }
t ^{\Gamma \setminus \gamma_P^{(i)}}
\cdot \delta^{\gamma_P^{(i)}}. \label{oweh}
\end{eqnarray}

We now assume that $t$ are the regularized but
unrenormalized distributions
and $\hat{t}$ are the regularized and renormalized distributions. Then
$\delta (t)$ describes the renormalization of the 
superficial divergence.

Taking only the parts of (\ref{oweh}) which are divergent when
the regularization is removed, we obtain:

\begin{eqnarray}
\delta^{\Gamma} =
-t ^{\Gamma}  -
\sum_{ P \in Part(J) \atop 1<|P|<n  }\delta^{\gamma_P^{(i)}}
t ^{\Gamma \setminus \gamma_P^{(i)}},
\end{eqnarray}
 
which has to be interpreted as the structure of singularities
in renormalization.
This recursive equation for the singularity of $\delta^{\Gamma}$
has exactly the same structure  as equation (\ref{antip}) for the
antipode (one has to identify a graph with the singularity of its 
numerical distribution).

\vspace{1cm}

{\bf Acknowledgement}: I want to thank Prof. K. Fredenhagen
for many useful discussions and a careful reading of the
manuscript.
The DFG is acknowledged for financial support.

\end{document}